\documentclass[a4paper,10pt]{article}
\usepackage{graphicx}
\usepackage{epstopdf}
\usepackage{amsfonts,amsmath,amssymb}
\usepackage{hyperref}

\usepackage{epsfig,multicol}

\newcommand\fverb{\setbox\pippobox=\hbox\bgroup\verb}
\newcommand\fverbit{\egroup\item[\fbox{\unhbox\pippobox}]}

\newbox\pippobox

\begin{document}
\title{\bf  Cosmic Acceleration From Matter--Curvature Coupling}
\author{Raziyeh Zaregonbadi\thanks{Electronic address: r$\_$zare@sbu.ac.ir}\, and\,
        Mehrdad Farhoudi\thanks{Electronic address: m-farhoudi@sbu.ac.ir}\,\,
\\
\small Department of Physics, Shahid Beheshti University, G.C.,
       Evin, Tehran 19839, Iran}
\date{\small September 2, 2016}
\maketitle
\begin{abstract}
\noindent
 We consider $f\left( {R,T} \right)$ modified theory of
gravity in which, in general, the gravitational Lagrangian is
given by an arbitrary function of the Ricci scalar and the trace
of the energy--momentum tensor. We indicate that in this type of
the theory, the coupling energy--momentum tensor is~not conserved.
However, we mainly focus on a particular model that matter is
minimally coupled to the geometry in the metric formalism and
wherein, its coupling energy--momentum tensor is also conserved.
We obtain the corresponding Raychaudhuri dynamical equation that
presents the evolution of the kinematic quantities. Then for the
chosen model, we derive the behavior of the deceleration
parameter, and show that the coupling term can lead to an
acceleration phase after the matter dominated phase. On the other
hand, the curvature of the universe corresponds with the deviation
from parallelism in the geodesic motion. Thus, we also scrutinize
the motion of the free test particles on their geodesics, and
derive the geodesic deviation equation in this modified theory to
study the accelerating universe within the spatially flat FLRW
background. Actually, this equation gives the relative
accelerations of adjacent particles as a measurable physical
quantity, and provides an elegant tool to investigate the timelike
and the null structures of spacetime geometries. Then, through the
null deviation vector, we find the observer area--distance as a
function of the redshift for the chosen model, and compare the
results with the corresponding results obtained in the literature.
\end{abstract}

\bigskip
\section{Introduction}\label{sec1}
\indent

The measurements of type Ia supernovae luminosity distances
indicate that the universe is currently undergoing an accelerated
expansion~\cite{1}--\cite{2}. The cause of this late time
acceleration has been one of the most challenging problems of
modern cosmology.  Such an acceleration requires, in general, some
components of negative pressure, with the equation of state
parameter, $w$, less than $ -1/3 $, to make an acceleration for
the universe. Several mechanisms, that are considered to be
responsible for this expansion, have been proposed, namely in
particular: the $\Lambda$CDM model, dark energy models and
modified gravities. In the $\Lambda$CDM model (e.g.,
Refs.~\cite{LCDM,3}), around $31\% $ of the mass--energy density
of the universe is made of the barionic and dark matter, and the
rest is constituted by the cosmological
constant~\cite{Ade-2013,Ade-2015}. Actually, the cosmological
constant is the simplest candidate which corresponds to the value
of $ w=-1 $, however, the value of the observed cosmological
constant is less than the Planck scale by a factor of $120$ orders
of magnitude. This problem, that is related to the
non--comparability of the value of dark energy density with the
field theoretical vacuum energy, is called the cosmological
constant problem, see, e.g., Refs.~\cite{3,Cos.pro1}--\cite{5}.

Another approach has been generalized by considering a source
term, with an equation of state parameter less than $ - 1/3$, that
is known as dark energy, see, e.g.,
Refs.~\cite{Peebles-2003}--\cite{Bahrehbakhsh-2013}. However, the
nature of dark energy is still unknown and actually, the nature of
both the dark matter and dark energy are one of the most important
issues in physics. In this approach, the observed cosmological
evidences imply on the existence of these components of the cosmic
energy budget, and even indicate that around $95\%$ of the
universe is composed by the dark matter and dark energy. Some
candidates for the dark matter (e.g., neutralinos and axions) and
dark energy (e.g., quintessence) have been considered, although no
direct detection has been reported until now. Also, various scalar
field models of dark energy have been studied in the literature,
e.g., Refs.~\cite{6}--\cite{Farajollahi-2012}.

On the other hand, the late time acceleration can be caused by
purely gravitational effects, i.e. one may consider modifying the
gravitational theory, see, e.g.,
Refs.~\cite{8}--\cite{Bahrehbakhsh-2011}. For instance, $f(R)$
gravities (as the simplest family of the higher--order gravities)
have been based on replacing the scalar curvature $ R $ in the
Einstein--Hilbert action with an arbitrary differentiable function
of it. In this regard, and as an example, it has been shown in
Refs.~\cite{chiba03,9} that by adding a term of $ 1/R $ to $ R $,
when the inverse curvature term dominates, one typically expects
that the universe expands with the desired late time acceleration,
however, this model suffers from a number of problems such as
matter instability, see, e.g., Ref.~\cite{Faraoni2006}. As an
another interesting example, it has been illustrated in
Ref.~\cite{Atazadeh2008} that with a $5$--dimensional $f(R)$ in
the brane world scenario (that converts to a scalar--tensor type
theory with a scalar field), an accelerated expanding universe
emerges for a suitable choice of the function $f$. However, $ f(R)
$ theories are somehow equivalent to the scalar--tensor theories,
in particular, to the specific type, namely the Brans--Dicke
theory, see, e.g., Ref.~\cite{Flanagan,Sotiriou2006} and
references therein. Actually, it has been
claimed~\cite{Sotiriou2006} that these theories are effectively
have the same phenomenology in most applications in cosmology and
astrophysics. For review on the generalized gravitational
theories, see, e.g., Refs.~\cite{10}--\cite{Clifton2012}.

In this work, we propose to probe the cosmological considerations
of another type of modified gravity theories, namely $ f(R,T)$
gravity, see, e.g., Refs.~\cite{Harko-2008}--\cite{shabani16b}.
Theory of $ f(R,T)$ gravity generalizes theories of gravity by
{\it a priori} incorporation of the trace of dustlike
matter\footnote{One already knows that the trace of the radiation
energy--momentum tensor is zero.}\
 energy--momentum tensor ($T$)
in addition to the Ricci scalar into the Lagrangian of geometry.
The reason for the {\it a priori} dependence on $ T $ may be the
inductions arising from some exotic imperfect fluids and/or
quantum effects (e.g., the conformal anomaly\footnote{See, e.g.,
Refs.~\cite{anomal1,anomal2}.}), and anyway, the $f(R,T)$ gravity
model depends on a source term, while somehow represents the
variation of the energy--momentum tensor with respect to the
metric. In another word, the \textit{a priori} appearance of the
matter in an unusual coupling with the curvature may also have
some relations with the known issues such as, geometrical
curvature induction of matter, a geometrical description of
physical forces, and a geometrical origin for the matter content
of the universe, see, e.g., Refs.~\cite{anomal2,fard} and
references therein. In this respect, whilst we first derive the
necessary equations for any arbitrary function of $f(R,T)$, the
main purpose of the work is to concentrate on a simplest and yet
the most plausible case of $f(R,T)$ gravity wherein the matter is
also minimally coupled to the curvature. Then, we obtain the state
parameter and the scale factor for this model in the matter
dominated and the acceleration phase of the universe. And mainly,
we investigate the effect of the cosmic matter and its coupling
with geometry in the evolution of the universe in addition to the
effect of pure geometry.

On the other side, to make our investigations for the acceleration
of the universe more instructive, we also scrutinize the motion of
the free test particles on their geodesics. In this regard, in the
context of relativistic gravitational theories, the curvature of
the spacetime plays a fundamental role instead of forces acting in
the Newtonian theory. On the other hand, the Einstein field
equations tell us how the curvature depends on the matter sources,
where one can attain the effects of the curvature in a spacetime
through the geodesic deviation equation ({\bf GDE}) via derivation
of the behavior of timelike, null and spacelike geodesics, as have
been performed in Refs.~\cite{32}--\cite{22}. Indeed, the GDE
gives the relative acceleration of two adjacent geodesics as a
measurable physical quantity, or a way that one can measure the
curvature of spacetime (analogous to the Lorentz force law), see
Refs.~\cite{Szekeres,mtw}. Moreover, it specifies the tendency of
free falling particles to recede or approach to each other while
moving under the influence of a varying gravitational field, i.e.
there exist internal effective tidal forces that cause the
trajectories of free particles to bend away or towards each other.
Also, the GDE provides a very elegant way of understanding the
structure of a spacetime and renders an invariant procedure of
characterizing the nature of gravitational
forces~\cite{Pirani1957}. And more interesting, this equation
contains many important results of standard cosmology such as the
Raychaudhuri dynamical equation~\cite{29}, the observer
area--distance (e.g., the Mattig relation for the dust
case~\cite{Mattig}), and how perturbations influence the
kinematics of null geodesics in resulting gravitational--lensing
effects~\cite{Clarkson} and references therein. In this respect,
as far as we are concerned, the GDE was studied within the
spatially flat Friedmann--Lema\^{\i}tre--Robertson--Walker ({\bf
FLRW}) spacetime geometries for general relativity plus
cosmological term in Ref.~\cite{22}. And thereafter, it has been
probed in the context of modified gravity theories, namely, the
Palatini formalism of $f(R)$~\cite{23}, the metric $f(R)$
gravity~\cite{24,25}, arbitrary matter--curvature coupling
theories~\cite{Harko2012}, a well--known class of $f(R)$
theory~\cite{26}, and $f(T)$ gravity (where in this theory, $T$ is
the torsion scalar arising from the torsion tensor)~\cite{27}.

The present work is organized as follows. The field equations and
the cosmological aspects of the modified $ f(R,T)$ gravity are
derived in Sect.~2, in where we specify the generalized form of
the Raychaudhuri dynamical equation that gives the evolution of
the kinematic quantities in the framework of modified $f(R,T)$
theory. The minimal coupling models are considered in Sect.~3,
wherein the matter dominated and the acceleration epochs
are discussed in a more detail for a particular model, as a simple
and plausible case of the minimal coupling models. Sect.~4 is
devoted to: obtain the GDE in this type of theory for timelike and
null geodesics within the spatially flat FLRW background, get a
particular case of the Raychaudhuri dynamical equation, and to
derive an useful relation measuring the observer area--distance as
a function of the redshift. Then in the same section, we attain
the results for the $\Lambda$CDM model and the particular minimal
coupling model, and compare them with each other and also with the
corresponding results of the Hu--Sawicki models~\cite{36} of
$f(R)$ theories obtained in Ref.~\cite{26} and $f(T)$ gravity
attained in Ref.~\cite{27}. Finally in Sect.~5, we present the
conclusions. Through the work, we use the sign convention $( { - ,
+ , + , + })$ and geometrical units with $c=1$.
\section{Modified Field Equations}\label{sec 2}
\indent

In this section, we obtain the field equations and some
corresponding dynamical parameters of $f(R,T)$ modified gravity in
four dimensional spacetime. The action is simply written in the
form
\begin{equation}\label{eq1}
S = \int {{d^4}x\sqrt { - g} \left[ {\frac{1}{{2\kappa
}}f\left({R,T} \right) + {L_m}} \right]},
\end{equation}
where $ \kappa  \equiv 8\pi G $, $g$ is the determinant of the
metric and $ {L_m} $  is the matter Lagrangian density, though for
simplicity and plausibly for our purposes, we assume it to be only
of the form of dustlike (non--relativistic) one. Also, the
energy--momentum tensor is usually defined as
\begin{equation}\label{eq2}
T_{\mu \nu }^{[m]} =  - \frac{2}{{\sqrt { - g} }}\frac{{\delta(
{\sqrt { - g} {L_m}})}}{{\delta {g^{\mu \nu }}}},
\end{equation}
where the lower case Greek indices run from zero to three and the
index $m$ stands for the dustlike matter. Variation of the action
with respect to the metric tensor gives the field equations
\begin{equation}\label{eq3}
{f_R}{R_{\mu \nu }} - \frac{f}{2}{g_{\mu \nu }} + \left({g_{\mu
\nu }{\Box}}- {\nabla _\mu }{\nabla _\nu }\right){f_R} = \left(
{\kappa + {f_T}} \right)T_{\mu \nu }^{[m]},
\end{equation}
where $\Box \equiv {\nabla_\mu }{\nabla^\mu }$ and we have defined
the following functions for the derivative of the function
$f(R,T)$ with respect to its arguments, i.e.
\begin{equation}\label{eq4}
{f_R} \equiv \frac{{\partial f(R,T)}}{{\partial R}}
\qquad\qquad\quad {\rm and} \qquad\qquad\quad {f_T} \equiv
\frac{{\partial f(R,T)}}{{\partial T}}.
\end{equation}
By contracting the field equations (\ref{eq3}), we attain the
scalar equation
\begin{equation}\label{eq5}
{f_R}R - 2f +3\, {\Box}{f_R}=\left( {\kappa  + {f_T}} \right)T.
\end{equation}

It is more instructive to write the field equations in the form of
the Einstein equations with an effective energy--momentum tensor,
as
\begin{equation}\label{eq6}
{G_{\mu \nu }} = \frac{\kappa }{{{f_R}}}\left( T_{\mu \nu }^{[m]}+
T_{\mu \nu }^{{\mathop{[\rm int]}} } \right) = {\kappa ^{[\rm
eff]}}T_{\mu \nu }^{[\rm eff]},
\end{equation}
where ${\kappa^{[\rm eff]}} \equiv \kappa /{f_R}$ and $T_{\mu \nu
}^{[\rm eff]} \equiv T_{\mu \nu }^{[m]} + T_{\mu \nu
}^{[{\mathop{\rm int}} ]}$. Also, the interaction/coupling
energy--momentum tensor has been defined as
\begin{equation}\label{eq7}
T_{\mu \nu }^{{\mathop{[\rm int]}} } \equiv \frac{1}{\kappa
}\left[{{f_T}T_{\mu \nu }^{[m]} + \frac{1}{2}(f - R{f_R}){g_{\mu
\nu }} + \left( {{\nabla _\mu }{\nabla _\nu } - {g_{\mu \nu
}{\Box}}} \right){f_R}} \right],
\end{equation}
that may be interpreted as a fluid composed from the interaction
between the matter and the curvature terms. Regarding the Bianchi
identity, from Eq. (\ref{eq6}), one easily achieves
\begin{equation}\label{eq conserv2}
{\nabla^\mu }T_{\mu\nu}^{[\operatorname{int}]} = \frac{{\left(
{{\nabla^\mu }{f_R}} \right)}}{\kappa }{G_{\mu \nu }},
\end{equation}
where the energy--momentum conservation of the dustlike matter,
i.e. ${\nabla^\mu }T_{\mu \nu }^{[m]} = 0$, and ${{f_R}\neq 0}$
have been used. From Eq. (\ref{eq conserv2}), it is obvious that
if $f_R$ is~not a constant, then the interaction energy--momentum
tensor will~not be conserved, i.e. some kind of internal force is
present. Hence, one can expect its effect to occur in the GDE.

Now, let us consider the spatially flat FLRW background with the
line element
\begin{equation}\label{eq8}
d{s^2} =  - d{t^2} + {a^2}\left( t \right)\left( {d{x^2} + d{y^2}
+ d{z^2}}\right),
\end{equation}
where $a\left(t \right)$ is the scale factor. Then, we show that
even if $f_R$ being a constant (although its corresponding
interaction energy--momentum tensor is conserved), due to ${f_T}
\neq 0 $, the related model can lead to an acceleration phase, and
causes some new terms to appear into the corresponding GDE. In
addition, a detailed calculations on Eq. (\ref{eq conserv2}) (that
has been furnished in the appendix) indicates that the following
constraint must hold for the FLRW metrics~\cite{20}
\begin{equation}\label{eq10}
\frac{3}{2}H{f_T} = {\dot f_T},
\end{equation}
where dot represents derivation with respect to the time and
$H\left( t \right) \equiv \dot a/a$  is the Hubble parameter. This
relation poses a restriction on the functionality of $f(R,T)$ for
these types of metrics.

Furthermore, in order to study the cosmological parameters, that
can be determined via the observational data, it is important to
derive some of the observational quantities which govern the
expansion history of the model. For instance, the kinematics of
the universe is described by the Hubble parameter and the
deceleration parameter, $q$. To have some expressions for these
parameters, let us also obtain the following equations from the
field equations, i.e.
\begin{equation}\label{eq11}
{H^2} = \frac{1}{{3{f_R}}}\left[ {\left( {\kappa  + {f_T}}
\right){\rho^{[m]}} + \frac{1}{2}\left( {R{f_R} - f} \right) -
3H{{\dot f}_R}} \right],
\end{equation}
as the Friedmann--like equation~\cite{28}, and
\begin{equation}\label{eq12}
3 {H^2} +2\dot H = - \frac{1}{{f_R}}\left[ {\frac{1}{2}( {f -
R{f_R}}) + 2H{{\dot f}_R} + {{\ddot f}_R}}\right],
\end{equation}
as the generalized Raychaudhuri equation. Besides, we can rewrite
Eq. (\ref{eq12}) by employing Eq. (\ref{eq5}) as
\begin{equation}\label{eq13}
\frac{{\ddot a}}{a} =  - \frac{1}{{3{f_R}}}\left[ {\left( {\kappa
+ {f_T}} \right){\rho ^{[m]}} - \frac{f}{2} - 3H{{\dot f}_R}}
\right].
\end{equation}
In deriving these relations, we have used the trace of matter
energy--momentum tensor, $ {T} =  - {\rho ^{[m]}} $, and assumed
that the universe is isotropic and homogeneous, thus only the time
derivatives do~not vanish, and hence, ${\Box}{f_R} =  - {\ddot
f_R} - 3H{\dot f_R} $.

On the other hand, we consider the interaction term as a perfect
fluid, and thus, the energy density and pressure of this fluid are
\begin{equation}\label{eq14}
{\rho ^{{\mathop{[\rm int]}} }} = - {g^{00}}T_{00}^{{\mathop{[\rm
int]}} } = \frac{1}{\kappa }\left[ {{f_T}{\rho ^{[m]}} +
\frac{1}{2}\left( {R{f_R} - f} \right) - 3H{{\dot f}_R}} \right]
\end{equation}
and
\begin{equation}\label{eq15}
{p^{{\mathop{[\rm int]}} }}
=\frac{1}{3}{g^{ii}}T_{ii}^{{\mathop{[\rm int]}} }=
\frac{1}{\kappa }\left[ {\frac{1}{2}\left( {f - R{f_R}} \right) +
2H{{\dot f}_R}}+{{\ddot f}_R} \right],
\end{equation}
where the lower case Latin indices run from one to three as the
spatial coordinates. And, using Eqs. (\ref{eq14}) and
(\ref{eq15}), we can rewrite Eqs. (\ref{eq11}) and (\ref{eq13}) as
\begin{equation}\label{eq16}
{H^2} = \frac{\kappa }{{3{f_R}}}\left( {{\rho^{[m]} } +
{\rho^{{\mathop{[\rm int]}} }}} \right) = \frac{1}{3}\kappa^{[\rm
eff]}{\rho^{[\rm eff]}}
\end{equation}
and
\begin{equation}\label{eq17}
\frac{{\ddot a}}{a} =  - \frac{\kappa }{{6{f_R}}}\left(
{{\rho^{[m]}} + {\rho^{[{\mathop{\rm int}} ]}} +
3{p^{[{\mathop{\rm int}} ]}}} \right) = - \frac{1}{6}\kappa^{[\rm
eff]}\left( {{\rho ^{[\rm eff]}} + 3{p^{[\rm eff]}}} \right),
\end{equation}
where $ {\rho^{[\rm eff]}} \equiv {\rho^{[m]} }
+{\rho^{{\mathop{[\rm int]}} }} $ and $ {p^{[\rm eff]}} \equiv
{p^{[m]}} + {p^{[{\mathop{\rm int}} ]}} $, however, in our case
${p^{[m]}}=0 $. Then, by inserting the corresponding definition of
the critical density at the present time, $\rho^{[{\rm
crit}]}(t_0) \equiv 3H_0^2/{\kappa ^{[ {\rm eff}]}}$, Eq.
(\ref{eq16}) reads
\begin{equation}\label{eq80}
{H^2} = H_0^2\left( {{\Omega ^{[ m]}} + {\Omega ^{[{{\mathop{\rm
int}}}]}}} \right),
\end{equation}
as usual in terms of the dimensionless density parameters,
${\Omega ^{[ i ]}} \equiv {\rho ^{[ i ]}}/\rho^{[{\rm
crit}]}(t_0)$. Also, the deceleration parameter is a dimensionless
measure of the cosmic acceleration of the expansion of the
universe, and is defined as $q\equiv -{\ddot a}a/{\dot
a}^2=-{\ddot a}H^{-2}/a$, where obviously if $ q < 0 $, then the
expansion of the universe will be an accelerated one, and if $ q >
0 $, it will describe a decelerating evolution. Inserting Eqs.
(\ref{eq16}) and (\ref{eq17}) into the definition of $q$, leads to
\begin{equation}\label{eq20}
q = 1 - \frac{1}{2}\left(\frac{R}{\kappa^{[\rm eff]}\rho^{[\rm
eff]}} \right).
\end{equation}
Incidentally, the state parameter in the equation of state (EoS),
for the modified gravitational theory is usually defined as
\begin{equation}\label{eq21}
{w^{[\rm eff]}} \equiv \frac{{{p^{[\rm eff]}}}}{{{\rho ^{[\rm eff]}}}}.
\end{equation}

In the next section, we investigate the cosmological
considerations of the universe in the matter dominated and the
 acceleration epochs for a particular minimal coupling
model of $ f(R,T) $ gravity.
\section{Minimal Coupling Model}\label{sec 3}
\indent

In the following, we consider a modified gravitational theory
where the matter is minimally coupled to the geometry in the type
of $ f\left( {R,T} \right) = g\left( R \right) + h\left( T \right)
$. However, the only form that can respect the conservation law of
the energy--momentum tensor, Eq. (\ref{eq10}) for the FLRW
metrics, must be in the form of $ h(T) \propto \sqrt { - T} $.
Hence, and also for later on convenience, we choose and write a
particular minimal coupling model as
\begin{equation}\label{eq22}
 f\left( {R,T} \right) = R - \alpha \sqrt { - T} ,
\end{equation}
where $\alpha $ is an adjustable
constant\rlap,\footnote{Obviously, $\alpha =0 $ reminds the
general relativity, and in the model, the $\alpha \neq 0$ case
corresponds to $f_T \neq 0 $.}\
 and thus, we have
\begin{equation}\label{eq23}
{f_R} = 1 \qquad\qquad\quad {\rm  and} \qquad\qquad\quad  {f_T} =
\frac{\alpha }{{2\sqrt { - T} }}.
\end{equation}
Applying Eq. (\ref{eq conserv2}) for the model, makes its
corresponding interaction energy--momentum tensor be conserved. In
addition, from Eq. (\ref{eq16}), one can easily derive the
behavior of the scale factor with respect to the comoving time for
the model as
\begin{equation}\label{eq18}
a\left( t \right) \propto {t^{2/\left[ 3\left( 1 + w^{\rm
[eff]}\right) \right]}}.
\end{equation}
Also, from Eqs. (\ref{eq14}) and (\ref{eq15}), the energy and
pressure densities for the coupling term, due to ${f_T} \neq 0 $,
are
\begin{equation}\label{eq24}
{\rho^{[{\rm int}]}} = \frac{\alpha }{\kappa }\sqrt { - T}=
\frac{\alpha }{\kappa }\sqrt {\rho^{[m]}}
\end{equation}
and
\begin{equation}\label{eq25}
{p^{[{\rm int}]}} =  - \frac{\alpha }{{2\kappa }}\sqrt {- T}  = -
\frac{1}{2}{\rho^{[{\rm int}]}}.
\end{equation}
Obviously, when $ \alpha  > 0$, one obtains  $ {\rho
^{{\mathop{[\rm int]}} }}> 0 $, that is more acceptable for the
interaction term as a perfect fluid. In this case, the pressure
from the interaction between the matter and geometry is always
negative.

On the other hand, from Eq. (\ref{eq5}), the Ricci scalar for the
particular minimal coupling model is
\begin{equation}\label{eq26}
R = \kappa {\rho^{[m]} } + \frac{5}{2}\alpha \sqrt{{\rho^{[m]}}}.
\end{equation}
Then, by inserting the Ricci scalar from the spatially flat FLRW
background into Eq. (\ref{eq26}), one obtains the coupling energy
density in terms of the Hubble parameter as
\begin{equation}\label{eq27}
{\rho^{[{\rm int}]}} = \frac{2}{\kappa }\left( {3{H^2} + 2\dot H}
\right).
\end{equation}
However, it is more instructive to define a dimensionless
parameter $x \equiv \alpha /( {\kappa \sqrt {\rho^{[m]}}})$, that,
in this model, reads $x=\rho^{[\rm int]}/\rho^{[m]}$. On the other
side, from the conservation law of the dustlike matter
energy--momentum tensor, one has $ {\rho^{[m]}} \propto {a^{ - 3}}
$, and hence, in terms of the redshift, we have
$\rho^{[m]}(z)=\rho^{[m]}_0(1+z)^3$, and consequently,
$x(z)=x(0)/(1+z)^{3/2}$ up to the present. Thus, we can rewrite
Eq. (\ref{eq80}) as
\begin{equation}\label{eq81}
{H^2(z)} = H_0^2\Omega _0^{[m]}\left(1 + z\right)^3\left(1 +
\frac{x(0)}{\left(1 + z\right)^{3/2}}\right).
\end{equation}
Also, from Eq. (\ref{eq20}), the deceleration parameter for this
model is
\begin{equation}\label{eq28}
q = \frac{1}{2} - \frac{3x}{4\left(1+x\right)}= \frac{1}{2} -
\frac{3x(0)}{4\left[(1+z)^{3/2}+x(0)\right]},
\end{equation}
and, from Eq. (\ref{eq21}), the state parameter reads
\begin{equation}\label{eq29}
{w^{[{\rm eff}]}} =
-\frac{x}{2\left(1+x\right)}=-\frac{x(0)}{2\left[(1+z)^{3/2}+x(0)\right]},
\end{equation}
that in turn indicates $q$ in terms of $w^{[{\rm eff}]}$ to be
$q=\left(1+3w^{[{\rm eff}]}\right)/2$.

Now, we investigate the evolution of the universe in the matter
dominated and the late time acceleration in the following
subsections, wherein, by using the fact that as the dustlike
density reduces by the time, within our proposal, we can compare
its amount with the adjustable constant $ {\left( {\alpha /\kappa
} \right)^2}$ as a criterion. Henceforth, we will illustrate that
in the matter dominated epoch, when $ {\rho^{\left[ m \right]}} $
is very larger than ${\left( {\alpha /\kappa } \right)^2} $, the
universe has a decelerated evolution. However, with the passage of
the time, the density of the matter reduces, and when $
{\rho^{\left[ m \right]}} = {\left( {\alpha /2\kappa } \right)^2}
$, the universe has a phase transition from the deceleration epoch
to the acceleration one. Furthermore, when the coupling energy
density dominates on the dustlike matter density, the equation of
state parameter is less than $ -1/3 $, that gives an accelerated
phase.
\subsection{Matter Dominated Phase}
\indent

By the assumption ${\rho ^{\left[ m \right]}} \gg {\left({\alpha
/\kappa } \right)^2} $, Eqs. (\ref{eq28}) and (\ref{eq29}) leads
to
\begin{equation}\label{eq30}
q \simeq \frac{1}{2}
\end{equation}
and
\begin{equation}\label{eq31}
w^{[\rm eff]} \simeq -\frac{{\alpha }}{{2\kappa \sqrt {{\rho
^{\left[ m \right]}}} }}\rightarrow 0^-.
\end{equation}
However, from Eqs. (\ref{eq24}) and (\ref{eq25}) (or similarly
from Eqs. (\ref{eq27}), (\ref{eq12}) and (\ref{eq15}) for the
model), we get $p^{[\rm int]} =-\alpha\sqrt
{\rho^{[m]}}/(2\kappa)$, although still in this assumption, one
can neglect $\rho^{[\rm int]}$ compared to $\rho^{[m]}$, and lets
$\rho^{[\rm eff]}\approx\rho^{[m]}$. That is, this case
corresponds to the matter dominated epoch. Also, by ${\rho^{[m]}}
\propto {a^{ - 3}}$, the time evolution of the scale factor, Eq.
(\ref{eq18}), approximately yields
\begin{equation}\label{eq33}
a\left( t \right) \propto {t^{2/3}},
\end{equation}
and hence, the density of the dustlike matter approximately is
${\rho ^{[m]}} \propto {t^{ - 2}} $, as expected.

On the other hand, the transition point from the deceleration to
the acceleration phase is when $ {\rho^{\left[ m \right]}} =
{\left( {\alpha /2\kappa } \right)^2}$, i.e., when
${\rho^{{\mathop{[\rm int]}} }} = 2{\rho ^{[m]}}$. Because at this
point, from Eq. (\ref{eq28}), the deceleration parameter vanishes,
that gives the transition redshift to be $z_{\rm
trans.}=\left[x(0)/2\right]^{2/3}-1$. However, with the parameter
value $x(0) = \rho_0^{[\rm int]}/\rho_0^{[m]} \simeq 7/3$, chosen
as a rough estimation, it results $z_{\rm trans.} \simeq 0.11 $,
that is less than the corresponding value of the $\Lambda $CDM
model. Also at this point, from Eq. (\ref{eq29}), the state
parameter reads
\begin{equation}\label{eq35}
{w^{[\rm eff]}} = - \frac{1}{3},
\end{equation}
that is, the universe has a transition from the deceleration era
to the acceleration one.
\subsection{Cosmic Acceleration Phase}
\indent

By reducing the matter density in the time, we consider when
${\rho^{\left[ m \right]}} \ll {\left( {\alpha /\kappa }
\right)^2} $, in which one can neglect $\rho^{[m]}$ compared to
$\rho^{[\rm int]}$, and lets $\rho^{[\rm eff]}\approx\rho^{[\rm
int]}$. In this phase of the universe, again due to ${f_T}\neq 0$,
the deceleration parameter, Eq. (\ref{eq28}), yields
\begin{equation}\label{eq36}
q \simeq  - \frac{1}{4}
\end{equation}
and the state parameter, Eq. (\ref{eq29}), leads to
\begin{equation}\label{eq37}
{w^{[\rm eff]}}  \simeq \frac{{ {p ^{[\rm int]} }}}{{{\rho ^{[\rm
int]} }}} = - \frac{1}{2},
\end{equation}
that is equal with the same value obtained for the dark energy
dominated era in Refs.~\cite{20,21}. Also, the time evolution of
the scale factor, Eq. (\ref{eq18}), approximately gives
\begin{equation}\label{eq39}
a\left( t \right) \propto {t^{\ 4/3}},
\end{equation}
hence,
the evolution of the Hubble parameter also approximately is
\begin{equation}\label{eq40}
H\left( t \right) \approx \frac{4}{{3t}}.
\end{equation}

Thus, in this phase of the universe, one approximately has $ {\rho
^{[m]}} \propto {t^{ - 4}}$ while ${\rho ^{{\mathop{[\rm int]}} }}
\propto {t^{ - 2}} $, and hence, the density of the interaction
between the matter and geometry dominates on the density of the
dustlike matter in the late time. Therefore, the coupling
energy--momentum tensor, due to ${f_T} \neq 0 $, can explain an
acceleration phase after the matter dominated phase of the
universe with no need of the mysterious dark energy. The evolution
of the deceleration parameter has been parameterically plotted
with respect to the dimensionless parameter $x$ in the left
diagram of Fig.~$1$, and with respect to the redshift (up to the
present) in the right diagram of it. The left diagram illustrates
again that: $x\rightarrow 0$ corresponds to the matter dominated
phase, also when $ {\rho ^{{\mathop{[\rm int]}} }} = 2{\rho ^{[m]}
}$, $q $ is zero and the transition from the deceleration to the
acceleration epoch occurs, and $x\rightarrow\infty$ relates to the
late time phase. In addition, the right diagram shows that the
deceleration parameter at z=0 is $q(0)= - 0.025$, i.e. an
accelerating universe.
\begin{figure}[h]
\begin{center}
\includegraphics[scale=0.6]{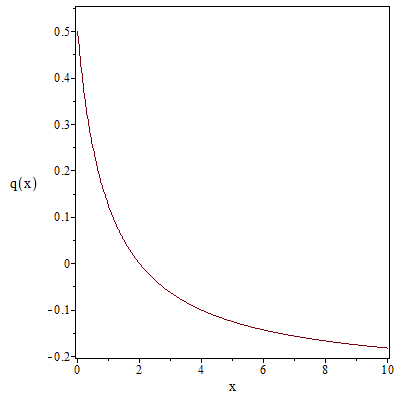}
\includegraphics[scale=0.6]{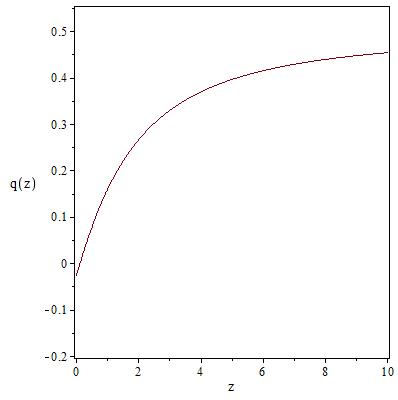}
\caption{The evolution of the deceleration parameter is depicted
for the particular minimal coupling model. The left diagram is
with respect to the dimensionless parameter $ x $, and the right
diagram is with respect to the redshift with the parameter value
$x(0)\simeq 7/3$ as a rough estimation.}
\end{center}
\end{figure}

\section{Geodesic Deviation Equation}\label{sec 4}
\indent

As mentioned, in the Einstein theory of gravitation, the physical
significance of the Riemann tensor is demonstrated by the geodesic
deviation concept. In this section, to investigate the relation
between the nearby geodesics and how the flux of geodesics
expands, we derive the GDE and the corresponding Raychaudhuri
dynamical equation in the context of $f(R,T) $ modified gravity.
Then, we mainly compare the results obtained for the particular
minimal coupling model with the results of the previous section in
order to have a better view on the cosmic acceleration phase. For
this purpose, let us consider the general expression for the
GDE~\cite{30,Inverno}
\begin{equation}\label{eq41}
\frac{{{D^2}{\eta ^\mu }}}{{D{\nu ^2}}} + R^{\mu}{}_{\nu \alpha
\beta } {V^\nu }{V^\alpha }{\eta^\beta }=0,
\end{equation}
where the parametric equation is given by $ {x^\mu }\left({\nu ,s}
\right)$, wherein $\nu$ is an affine parameter along the geodesics
and $s$ labels distinct geodesics. Also, ${\eta ^\mu} =
d{x^\mu}/ds $ is the orthogonal deviation vector of two adjacent
geodesics, the normalized vector field ${V^\mu} = d{x^\mu}/d\nu $
is tangent to the geodesics and $ {D}/{D\nu } $ is the covariant
derivative along the curve.

Now, to obtain the relation between the geometrical properties of
the spacetime with the field equations governed from the modified
gravitational theory, one can employ the well--known relation
\begin{equation}\label{eq42}
{R_{\mu \nu \alpha \beta}}={C_{\mu \nu \alpha \beta}}
+\frac{1}{2}\left( {{g_{\mu \alpha}}{R_{\nu \beta}} - {g_{\mu
\beta}}{R_{\nu \alpha}} + {g_{\nu \beta}}{R_{\mu \alpha}} -
{g_{\nu \alpha}}{R_{\mu \beta}}} \right) - \frac{1}{6}{R}\left(
{{g_{\mu \alpha}}{g_{\nu \beta}} - {g_{\mu \beta}}{g_{\nu
\alpha}}} \right),
\end{equation}
in $4$--dimensions. However, for the conformally flat spacetime,
the Weyl tensor  $ {C_{\mu \nu \alpha \beta}} $  is identically
zero. Then, by assumption ${f_R} \neq 0 $, and inserting the Ricci
tensor from the field equations (\ref{eq3}) and the Ricci scalar
from Eq. (\ref{eq5}) into relation (\ref{eq42}), we obtain
\begin{equation}\label{eq43}
\begin{array}{l}
{R_{\mu \nu \alpha \beta}}= \frac{1}{{2{f_R}}} \bigg[ \left(
{\kappa + {f_T}} \right)\left( {g_{\mu \alpha}}T_{\nu \beta}^{[m]}
- {g_{\mu \beta}}T_{\nu \alpha}^{[m]} + {g_{\nu \beta}}T_{\mu
\alpha}^{[m]} - {g_{\nu \alpha}}T_{\mu \beta}^{[m]} \right) +
f\left( {g_{\mu \alpha}}
{g_{\nu \beta}} - {g_{\mu \beta}}{g_{\nu \alpha}} \right)\bigg]  \\
\qquad\qquad + \frac{1}{{2{f_R}}} \left( {g_{\mu \alpha}}{D_{\nu
\beta}} - {g_{\mu \beta}}{D_{\nu \alpha}}
+ {g_{\nu \beta}}{D_{\mu \alpha}} - {g_{\nu \alpha}}{D_{\mu \beta}} \right){f_R}\\
\qquad\qquad - \frac{1}{{6{f_R}}}\bigg[ {\left( {\kappa + {f_T}}
\right)T + 2f - 3\, {\Box}{f_R}} \bigg]\left( {{g_{\mu
\alpha}}{g_{\nu \beta}} - {g_{\mu \beta}}{g_{\nu \alpha}}}
\right),
\end{array}
\end{equation}
where ${D_{\mu \nu}} \equiv {\nabla _\mu}{\nabla _\nu}- {g_{\mu
\nu}{\Box}}$. Also, as assumed, for the dustlike matter flow as a
perfect fluid, the energy--momentum tensor obviously is $T_{\mu
\nu }^{[m]} ={\rho^{[m]} }\, {u_\mu }u_\nu$, where $ {u^\mu } $ is
the comoving velocity vector to the matter flow with $ {u_\mu
}{u^\mu } =  - 1 $, and hence, we get
\begin{equation}\label{eq45}
\begin{array}{l}
-R^{\mu}{}_{\nu \alpha \beta } {V^\nu }{V^\alpha }{\eta^\beta } = \\
\qquad\frac{1}{{2{f_R}}}\left[ \left( {\kappa  + {f_T}} \right)
{{\rho^{[m]}}\left( {\delta _\alpha ^\mu {u_\nu }{u_\beta } -
\delta _\beta ^\mu {u_\nu }{u_\alpha } + {g_{\nu \beta }}{u^\mu }
{u_\alpha } - {g_{\nu \alpha }}{u^\mu }{u_\beta }} \right) +
f\left( {\delta _\alpha ^\mu {g_{\nu \beta }} - \delta _\beta ^\mu
{g_{\nu \alpha }}} \right)} \right]{V^\nu }{V^\alpha }{\eta^\beta }\\
\qquad+ \frac{1}{{2{f_R}}}\left[ {\left( {\delta _\alpha ^\mu
{D_{\nu \beta }} - \delta _\beta ^\mu {D_{\nu \alpha }} + {g_{\nu
\beta }}D^{\mu}{}_\alpha   - {g_{\nu \alpha }}
D^{\mu}{}_\beta } \right){f_R}} \right]{V^\nu }{V^\alpha }{\eta^\beta }\\
\qquad- \frac{1}{{6{f_R}}}\left[ { - \left( {\kappa  +
{f_T}}\right){\rho ^{[m]}} + 2f - 3\, {\Box}{f_R}} \right]\left(
{\delta _\alpha ^\mu {g_{\nu \beta }} - \delta _\beta ^\mu {g_{\nu
\alpha }}} \right){V^\nu }{V^\alpha }{\eta^\beta }.
\end{array}
\end{equation}
By substituting the total energy $ E =  - {V_\mu}{u^\mu} $,
$\varepsilon\equiv {V^\mu}{V_\mu}$, the relations ${\eta
_\mu}{u^\mu} =0={\eta _\mu}{V^\mu} $ (note that, in the comoving
frame, we have $ {\eta ^0}=0 $), $ {D_{00}}{f_R} =  - 3H{\dot f_R}
$ and ${D_{ij}}{f_R} = {g_{ij}}\left( {{{\ddot f}_R} + 2H{{\dot
f}_R}} \right) $ into Eq. (\ref{eq45}), the ``force term" for the
geodesic congruences yields
\begin{equation}\label{eq46}
R^{\mu}{}_{\nu \alpha \beta } {V^\nu }{V^\alpha }{\eta^\beta }=
\frac{-1}{{2{f_R}}}\left\{ {\left[\left( {\kappa  + {f_T}} \right)
{{\rho ^{[m]}}+ {{\ddot f}_R} - H{{\dot f}_R}} \right]{E^2} +
\frac{1 }{3}\left[ {{\rho ^{[m]}}\left( {\kappa  + {f_T}} \right)
+ f + 3{{\ddot f}_R} + 3H{{\dot f}_R}}\right]{\varepsilon} }
\right\}{\eta ^\mu }.
\end{equation}
As it is obvious, even if $f_R$ being a constant, due to $f_T \neq
0$, still some new terms appear into the corresponding GDE of the
related model. Finally, by replacing Eqs. (\ref{eq14}) and
(\ref{eq15}) into Eq. (\ref{eq46}), we obtain
\begin{equation}\label{eq47}
R^{\mu}{} _{\nu \alpha \beta }{V^\nu }{V^\alpha }{\eta^\beta }=
-\frac{1}{2}\kappa^{[\rm eff]}\left[ {\left( {{\rho ^{[\rm eff]}}
+ { p^{[\rm eff]}}} \right){E^2} + \frac{1}{3}\left( {{\rho ^{[\rm
eff]}} + 3{ p^{[\rm eff]}} + \frac{R}{\kappa^{[\rm eff]}}}
\right)\varepsilon } \right]{\eta ^\mu}
\end{equation}
for the FLRW metrics, which is a generalization of the Pirani
equation~\cite{31}. Therefore, the GDE for the modified
gravitational theory is
\begin{equation}\label{eq48}
\frac{D^2\eta^\mu}{D\nu^2} + \frac{1}{2}\kappa^{[\rm eff]}\left[
{\left( {{\rho ^{[\rm eff]}} + { p^{[\rm eff]}}} \right){E^2} +
\frac{1}{3}\left( {{\rho ^{[\rm eff]}} + 3{ p^{[\rm eff]}} +
\frac{R}{\kappa^{[\rm eff]}}} \right)\varepsilon } \right]{\eta
^\mu}=0.
\end{equation}

From the recent equation, one can obviously attain the GDE for the
particular minimal coupling model and the $\Lambda$CDM model by
inserting the corresponding Ricci scalar of each one of these
models. In this regard, for the particular minimal coupling model,
by substituting Eqs. (\ref{eq24}), (\ref{eq25}) and (\ref{eq26})
into the GDE (\ref{eq48}), gives
\begin{equation}\label{eq49}
\frac{{{D^2}{\eta ^\mu}}}{{D{\nu ^2}}} + \frac{1}{2}\left[
{\left({\kappa {\rho ^{[m]}} + \frac{{\alpha \sqrt {{\rho ^{[m]}}}
}}{2}} \right){E^2} + \frac{2}{3}\left( {\kappa {\rho ^{[m]}} +
\alpha \sqrt {{\rho ^{[m]}}} } \right)\varepsilon } \right]{\eta
^\mu}=0.
\end{equation}
On the other hand, for the $ \Lambda$CDM model, i.e. $f(R) = R -
2\Lambda  $, from the corresponding Eq. (\ref{eq5}), we have
\begin{equation}\label{eq50}
R = \kappa {\rho ^{[m]}} + 4\Lambda,
\end{equation}
and Eqs. (\ref{eq14}) and (\ref{eq15}) result
\begin{equation}\label{eq51}
{\rho ^{[{\mathop{\Lambda}} ]}}=\frac{\Lambda }{\kappa }=
{-p^{[{\mathop{\Lambda}} ]}}.
\end{equation}
Thus, the GDE (\ref{eq48}) for this case reads
\begin{equation}\label{eq52}
\frac{D^2\eta^\mu}{D\nu^2} +
\frac{1}{2}\left[\left(\kappa\rho^{[m]}\right){E^2} +
\frac{2}{3}\left(\kappa\rho^{[m]} + \Lambda
\right)\varepsilon\right]{\eta^\mu }=0.
\end{equation}
The comparison of Eqs. (\ref{eq49}) and (\ref{eq52}) explicitly
indicates the effect of $f_T\neq 0$.

In the following, we evaluate the timelike and the null geodesics
in two subsections. However before we proceed, let us remind that,
as we expect, each of the ``force term" in these obtained GDEs is
proportional to ${\eta ^\mu }$ itself, and hence, only induces
isotropic deviation. That is, in the FLRW case, we only have such
a term which gives the spatial isotropy of spacetime, and the
deviation vector ${\eta ^\mu }$ only changes in magnitude but not
in direction~\cite{32,22}. Although, in an anisotropic universe,
like the Bianchi I, the corresponding GDE also reflects a change
in the direction of the deviation vector as shown in
Ref.~\cite{33}.
\subsection{GDE For Timelike Vector Fields}
\indent

For the timelike vector fields defined by the velocities of the
comoving observers, within the spatially flat FLRW background, one
has $\varepsilon = - 1$, $E = 1 $ and the affine parameter $ \nu $
matching with the proper time $ t $ of the comoving fundamental
observers up to a constant. Thus, the GDE (\ref{eq48}) reduces to
\begin{equation}\label{eq53}
\frac{{{D^2}{\eta ^\mu }}}{{D{t^2}}} + \frac{1}{6 {f_R}}\left(
{2\kappa {\rho ^{\left[ {\rm eff} \right]}} - R} \right){\eta ^\mu
}=0,
\end{equation}
and hence, for the particular minimal coupling model, it reads
\begin{equation}\label{eq54}
\frac{{{D^2}{\eta ^\mu }}}{{D{t ^2}}} + \frac{1}{6}\left( {\kappa
{\rho ^{[m]}} - \frac{{\alpha \sqrt {{\rho ^{[m]}}} }}{2}}
\right){\eta ^\mu }=0.
\end{equation}
However, the deviation vector can be written in terms of the
comoving tetrad frame (the Fermi--transported along unit tangent
to geodesic) as $ {\eta ^\mu} = a\left( t \right){e^\mu}$, such
that, it connects adjacent geodesics in the radial direction.
Then, as isotropy (with no shear and spatial rotation) implies
\begin{equation}\label{eq55}
\frac{{D{e^\mu}}}{{Dt}} = 0,
\end{equation}
therefore, Eq. (\ref{eq54}) yields
\begin{equation}\label{eq56}
\frac{{\ddot a}}{a} = - \frac{\kappa }{6}\left( {{\rho ^{\left[
m\right]}} - \frac{{{\rho ^{\left[ {{\mathop{\rm int}} }
\right]}}}}{2}} \right).
\end{equation}
This equation is a particular case of the so called Raychaudhuri
dynamical equation. Obviously, Eq. (\ref{eq56}) indicates that if
the dustlike matter dominates, the universe will decelerate and,
when $ {\rho ^{{\mathop{[\rm int]}} }} = 2{\rho ^{[m]}} $, the
acceleration for timelike geodesics vanishes. Also, in the late
time universe (when  $ {\rho ^{[\rm int ]}} \gg {\rho ^{[m]}} $),
the acceleration occurs. These results are consistent with the
results obtained in Sect.~3.

On the other hand, the GDE, for the timelike vector fields of the
comoving observers in the $\Lambda $CDM model, gives
\begin{equation}\label{eq57}
\frac{{\ddot a}}{a} =  - \frac{1}{6}\left( {\kappa {\rho^{[m]}} -
2\Lambda } \right),
\end{equation}
and when ${\rho ^{[m]}} = 2{\rho ^{\left[ \Lambda  \right]}} $,
the universe transition from the deceleration to the acceleration
phase occurs. Also when ${\rho^{[\Lambda ]}} \gg {\rho^{[m]}}$, we
have an accelerated universe in the late time.
\subsection{GDE For Null Vector Fields}
\indent

In this subsection, we investigate the past--directed null vector
fields, where in this case, still using the spatially flat FLRW
background, one has $\varepsilon = 0$. By considering $ {\eta ^\mu
} = \eta {e^\mu } $ and using a parallelly propagated and aligned
coordinate basis, i.e. ${D{e^\mu }/D\nu } = {V^\alpha }{\nabla
_\alpha }{e^\mu } = 0 $, the GDE (\ref{eq48}) reduces to
\begin{equation}\label{eq58}
\frac{{{d^2}\eta }}{{d{\nu ^2}}} + \frac{1}{2}\kappa^{[\rm
eff]}\left( { {\rho ^{[\rm eff]}} + {p^{[\rm eff]}}}
\right){E^2}\eta=0.
\end{equation}

At first, according to the general relativity (within the
spatially flat FLRW spacetimes) discussed in Ref.~\cite{22}, all
families of the past--directed null geodesics experience focusing
if provided the condition $\kappa \left( {\rho  + p} \right) > 0$.
In particular, for the cosmological constant case, in which the
corresponding equation of state is $ {p} =  - {\rho} $, there is
no focusing effect of the null geodesics. And hence, the solution
of equation (\ref{eq58}), in this particular case, is $ \eta
\left( \nu \right) = {C_1}\nu + {C_2} $ (where $ {C_1} $ and
${C_2} $ are integration constants) that is equivalent to the
Minkowski spacetime. However, in this regard, the corresponding
focusing condition for the $f(R,T) $ gravity model
is\footnote{Note that, the focusing of geodesics can also be
described by geometrical terms using the Raychaudhuri equation, in
particular, it only depends on the deceleration parameter and the
physical velocity of the geodesics, see Ref.~\cite{Albareti2012}.
Indeed, expressions like relation (\ref{eq59}) are known as energy
conditions and, as the Weyl tensor is zero for the FLRW metrics,
can be derived using the Raychaudhuri equation instead of the GDE,
see, e.g. for modified gravity,
Refs.~\cite{Santos2007,Albareti2013}.}
\begin{equation}\label{eq59}
{\kappa ^{[\rm eff]}}\left( {{\rho^{[\rm eff]}} + {p^{\left[ {\rm
eff} \right]}}} \right) > 0.
\end{equation}
Now, for the particular minimal coupling model, as the GDE
(\ref{eq58}) reads
\begin{equation}\label{eq60}
\frac{{{d^2}{\eta}}}{{d{\nu ^2}}} + \frac{1}{2}\left( {\kappa
{\rho ^{[m]}} + \frac{{\alpha \sqrt {{\rho ^{[m]}}} }}{2}}
\right){E^2}{\eta}=0,
\end{equation}
its focusing condition is actually provided when
\begin{equation}\label{eq61}
\left( {2{\rho ^{[m]}} + {\rho ^{\left[ {{\mathop{\rm int}} }
\right]}}} \right) > 0
\end{equation}
and, in turn, when $\sqrt {{\rho ^{[m]}}}>-\alpha/(2\kappa)$ are
satisfied, that is always confirmed for positive values of
$\alpha$.

In continuation, it is more appropriate to write the deviation
vector for the null geodesics as a function of the redshift rather
than the proper time. This can obviously be performed through the
transformation between the affine parameter and the redshift,
where we also rewrite the GDE with respect to it. For this
purpose, let us start by the simple differential operator
\begin{equation}\label{eq62}
\frac{d}{{d\nu }} = \frac{{dz}}{{d\nu }}\frac{d}{{dz}},
\end{equation}
that obviously yields
\begin{equation}\label{eq63}
\frac{d^2}{d\nu^2}=\left(\frac{dz}{d\nu}\right)^2\frac{d^2}{dz^2}+
\frac{d^2z}{d\nu^2}\frac{d}{dz}=\left(\frac{d\nu}{dz}\right)^{-2}\left[
\frac{d^2}{dz^2} - (\frac{d\nu}{dz})^{-
1}\frac{d^2\nu}{dz^2}\frac{d}{dz}\right].
\end{equation}
Also, for the null geodesics, we have the well--known relation
$(1+z) = a_0/a = E/E_0$, that by assuming ${a_0} = 1 $ (as the
present--day value of the scale factor for the flat spacetime) and
${E_0}$ as a constant, leads to
\begin{equation}\label{eq65}
\frac{{dz}}{{1 + z}} =  - \frac{{da}}{a} = \frac{{dE}}{E},
\end{equation}
and thus, one obtains
\begin{equation}\label{eq66}
dz =  - \left( {1 + z}\right)\frac{{\dot a}}{a}\frac{{dt}}{{d\nu
}}d\nu  =  - \left( {1 + z} \right)HEd\nu.
\end{equation}
However, by substituting for $E$, one can rewrite Eq. (\ref{eq66})
as
\begin{equation}\label{eq67}
\frac{{d\nu }}{{dz}} = - \frac{1}{{{E_0}H{{\left( {1 + z}
\right)}^2}}},
\end{equation}
and hence, gets
\begin{equation}\label{eq68}
\frac{{{d^2}\nu }}{{d{z^2}}} = \frac{1}{{{E_0}H{{\left({1 + z}
\right)}^3}}}\left[ {2 + \frac{{\left( {1 + z} \right)}}{H}\left(
{\frac{{dH}}{{dz}}} \right)} \right].
\end{equation}
Besides, we have
\begin{equation}\label{eq69}
\frac{{dH}}{{dz}} = \frac{{d\nu }}{{dz}}\frac{{dt}}{{d\nu
}}\frac{{dH}}{{dt}}=  - \frac{{\dot H}}{{H\left( {1 + z}
\right)}},
\end{equation}
that by replacing it into Eq. (\ref{eq68}), leads to
\begin{equation}\label{eq70}
\frac{{{d^2}\nu }}{{d{z^2}}} = \frac{1}{{{E_0}H{{\left( {1 +
z}\right)}^3}}}\left( {2 - \frac{{\dot H}}{{{H^2}}}} \right).
\end{equation}
On the other hand, from the definition of the Hubble parameter,
one obviously has $\dot H ={\ddot a}/a - {H^2}$, that by inserting
it into Eq. (\ref{eq70}), while using Eq. (\ref{eq17}), gives
\begin{equation}\label{eq72}
\frac{{{d^2}\nu }}{{d{z^2}}} = \frac{1}{{{E_0}H{{\left( {1 +
z}\right)}^3}}}\left[ {3 + \frac{{{\kappa ^{\left[ {\rm eff}
\right]}}\left( {{\rho ^{\left[ {\rm eff} \right]}} + 3{p^{\left[
{\rm eff} \right]}}} \right)}}{{6{H^2}}}} \right].
\end{equation}
Finally, by substituting Eqs. (\ref{eq67}) and (\ref{eq72}) into
Eq. (\ref{eq63}), we obtain
\begin{equation}\label{eq73}
\frac{{{d^2}\eta }}{{d{\nu ^2}}} = E_0^2{H^2}{\left( {1 +
z}\right)^4}\left\{ {\frac{{{d^2}\eta }}{{d{z^2}}} +
\frac{1}{{\left( {1 + z} \right)}}\left[ {3 + \frac{{{\kappa
^{\left[ {\rm eff} \right]}}\left( {{\rho ^{\left[ {\rm eff}
\right]}} + 3{p^{\left[ {\rm eff} \right]}}} \right)}}{{6{H^2}}}}
\right]\frac{{d\eta }}{{dz}}} \right\},
\end{equation}
where, by inserting Eq. (\ref{eq58}) into it, leads to the null
GDE corresponding to the modified gravity as
\begin{equation}\label{eq74}
\frac{{{d^2}\eta }}{{d{z^2}}} + \frac{3}{{1 + z}}\left[ {1
+\frac{{{\kappa ^{[\rm eff]}}}}{{18{H^2}}}\left( {{\rho ^{[\rm
eff]}} + 3{p^{[\rm eff]}}} \right)} \right]\frac{{d\eta }}{{dz}} +
\frac{{{\kappa ^{[\rm eff]}}}}{{2{{\left( {1 + z}
\right)}^2}{H^2}}}\left( {{\rho ^{[\rm eff]}} + {p^{[\rm eff]}}}
\right)\eta  = 0.
\end{equation}
Also, by substituting $ {\rho ^{[\rm eff]}} $ and $ {p^{[\rm
eff]}} $ from definitions (\ref{eq16}) and (\ref{eq21}) into Eq.
(\ref{eq74}), reduces the null GDE as
\begin{equation}\label{eq75}
\frac{{{d^2}\eta }}{{d{z^2}}} + \frac{{\left( {7 + 3{w^{[\rm
eff]}}}\right)}}{{2\left( {1 + z} \right)}}\frac{{d\eta }}{{dz}} +
\frac{{3\left( {1 + {w^{[\rm eff]}}} \right)}}{{2{{\left( {1 + z}
\right)}^2}}}\eta  = 0,
\end{equation}
where it obviously depends only on $ {w^{[\rm eff]}} $ as a
function of the redshift, i.e. as a function of the cosmological
evolution.

At last, by inserting Eq. (\ref{eq29}) into (\ref{eq75}), we can
obtain the GDE for the particular minimal coupling model as
\begin{equation}\label{eq76}
\frac{d^2\eta}{dz^2} + \frac{14 + 11x(z)}{4\left(1 +
z\right)\left(1 + x(z)\right)}\frac{d\eta}{dz} + \frac{3\left(2 +
x(z)\right)}{4\left(1 + z\right)^2\left(1 + x(z)\right)}\eta = 0,
\end{equation}
where again $x(z)=x(0)/(1+z)^{3/2}$ up to the present. This
equation has a $Q$--Legendre solution as
\begin{equation}
\eta \left( z\right) = \frac{{{C_1}}}{{1 + z}} +
\frac{{{C_2}}}{{{{\left( {1 + z} \right)}^{7/8}}}}Q_{-
1/6}^{1/6}\left( {\frac{{\sqrt {49 + \left( {21z + 21}
\right)\sqrt {1 + z} } }}{7}} \right),
\end{equation}
where $ C_1 $ and $ C_2 $ are integration constants related to
$x(0)$. As appropriate initial conditions, we set, at $ z=0 $, the
deviation vector to be zero and its first derivative to be $1$,
and then, plot the deviation vector with respect to the redshift
in the left diagram of Fig.~$2$, wherein the corresponding term of
the $\Lambda$CDM model has also been numerically drawn for
comparison.
\begin{figure}[h]
\begin{center}
\includegraphics[scale=0.55]{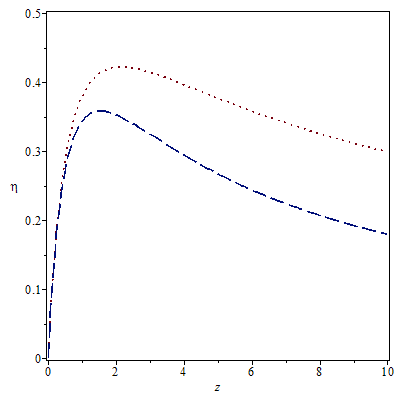}
\includegraphics[scale=0.55]{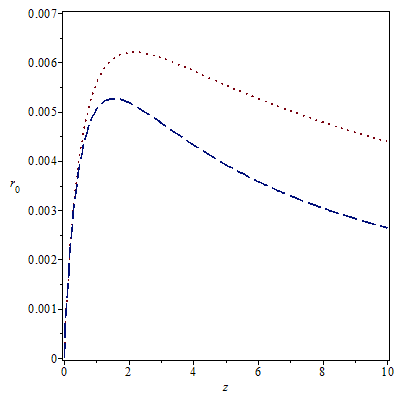}
\caption{The deviation vector (left) and the observer
area--distance (right) are depicted with respect to the redshift,
for the $\Lambda$CDM model (the dotted curve) and for the
particular minimal coupling model (the dashed curve), with the
appropriate initial conditions $\eta(z)|_{z=0}=0$ and
$d\eta(z)/dz|_{z=0}=1$, and the parametric values $H_0\simeq 68$
(km/s)/Mpc and $x(0)\simeq 7/3$ as rough estimations.}
\end{center}
\end{figure}

Now, we are in a position to indicate the observer area--distance,
${r_0} $, first derived by Mattig~\cite{Mattig}. At first, in a
spherically symmetric spacetime, for instance in the FLRW
universes, the magnitude of the deviation vector $ \eta $ is
related with the proper area $dA$ of a source in a redshift $ z $
as $ d\eta \propto \sqrt {dA} $, and hence, the observer
area--distance can be found as a function of the redshift in units
of the present--day Hubble radius $H_0^{ - 1} $, see, e.g.,
Ref.~\cite{34}. That is, by its definition, we get
\begin{equation}\label{eq77}
{r_0}\left( z \right) := \sqrt {\left| {\frac{{d{A_0}\left( z
\right)}}{{d\Omega}_0}}\right|}  = \left| {\frac{{\eta \left( {z'}
\right)\left| {_{z' = z}} \right.}}{{d\eta \left( {z'}
\right)/dl\left| {_{z' = 0}} \right.}}} \right|,
\end{equation}
where $ {A_0} $ is the area of the object, $\Omega  $ is the solid
angle and $ dl = a\left( t \right)dr $ with \textit{r} as the
comoving radial coordinate in the FLRW metrics. Using the fact
that $d/dl = H\left( {1 + z} \right)d/dz $ and choosing the
deviation vector to be zero at $z=0$, thus ${r_0} $ reads
\begin{equation}\label{eq78}
{r_0}\left( z \right) = \left| {\frac{{\eta \left( z
\right)}}{H(0){d\eta\left( {z'} \right)/dz'\left| {_{z' = 0}}
\right.}}} \right|,
\end{equation}
where practically $H(0) $ can be evaluated from the modified
Friedmann equation (\ref{eq81}) at $z=0 $. Analytical expression
for the observer area--distance in general relativity with no
cosmological constant obtained in Refs.~\cite{22,35}. However, in
our case, the plot of Eq. (\ref{eq78}) for the particular minimal
coupling model has been depicted in the right diagram of Fig.~$2$
while employing the solution of the deviation vector in the left
diagram of it with the approximate value of $68$ (km/s)/Mpc for
$H_0$~\cite{Ade-2015}.

The comparison of the diagrams of Fig.~$2$ with the corresponding
diagrams of the $ \Lambda$CDM model in Refs.~\cite{22,26}
illustrates that the general behavior of the null geodesic
deviation and the observer area--distance in the particular
minimal coupling model are almost similar to the $ \Lambda$CDM
model. Hence, this model can remain phenomenologically viable and
being tested with the observational data. Also, the comparison of
the results of the model, with the corresponding results of the
Hu--Sawicki models\footnote{It has been claimed~\cite{36} that the
Hu--Sawicki models of $f(R)$ theories (i.e., $f(R) = aR -
{m^2}{b{\left( {R/{m^2}} \right)^n}/\left[ 1 + c{\left( {R/{m^2}}
\right)^n} \right]}$, where $a$, $b$ and $c$ are dimensionless
constants and ${m^2}$ is related to the square of the Hubble
parameter) supply a viable cosmological evolution, while these
models have been studied in the range of astrophysical and
cosmological situations.}\
 (considered in Ref.~\cite{26}) and the $f(T)$ gravity (obtained in
Ref.~\cite{27}), shows, in general, similar behaviors for the null
geodesic deviation and the observer area--distance.

\section{Conclusions}\label{sec 5}
\indent

We have considered the $f(R,T)$ modified gravity, in particular,
have focused on the minimal coupling of the matter and geometry
within the homogeneous and isotropic spatially flat FLRW
background. Considering this particular case of the theory, by the
requirement of the conservation law of the energy--momentum
tensor, we choose the model to be a simple and the most plausible
of the form $f(R,T)=R - \alpha \sqrt { - T} $, with $ \alpha $ as
a positive and adjustable constant. Also, we assume that the
universe is made of the dustlike matter and a perfect fluid for
the interaction/coupling between the matter and geometry. Then, by
the fact that the density of the matter reduces by the time, we
compare it with the constant value of $\left(\alpha
/\kappa\right)^2$ as a criterion. Hence, we have shown that, the
case ${\rho^{[m]}} \gg \left(\alpha /\kappa\right)^2$ corresponds
to the matter dominated epoch with a decelerated evolution, and
the case ${\rho^{[m]}}\ll \left( \alpha /\kappa \right)^2$  is
related to an accelerated phase. To study the cosmology of the
model in the matter dominated and the acceleration phases, we have
obtained some cosmological parameters such as the state and the
deceleration parameters. We find that the value of the state
parameter in the coupling density dominated is less than $ -1/3 $
and the deceleration parameter attains a negative value, thus, an
accelerated universe arises by the coupling between the matter and
geometry. It has been indicated that, when $\rho^{[ m ]} = {(
{\alpha /2\kappa })^2}$, the deceleration parameter is zero, and
the universe has a transition from the deceleration phase to the
acceleration one. Also, we have shown that the coupling energy
density depends on the Hubble parameter, and thus, with the
passage of the time, it gets larger than the matter density.
Although we find that the minimal coupling model causes an
acceleration phase, but its obtained equation of state parameter
is less than the corresponding observed data. Nevertheless, if one
considers a more general model, yet in this context but with
non--conserved interaction energy--momentum tensor, then, one
would expect to get results more closer to the observations.

On the other hand, to make our investigations more instructive, we
have also scrutinized the motion of the free test particles on
their geodesics, and have obtained the GDE in context of the
$f(R,T)$ modified gravity, which provides an elegant tool to
investigate the structure of the spacetime, and to describe the
relative acceleration of the neighboring geodesics as a measurable
physical quantity. The coupling between the matter and curvature
leads to the appearance of the new internal ``force terms" in the
GDE. Moreover, through the corresponding GDE, we have derived the
generalized Pirani equation for the chosen model. Then, we have
investigated the timelike and the null vector fields in this type
of modified theory. The case of fundamental observers gives us the
generalized Raychaudhuri equation. Also, we have obtained the
observer area--distance for this case, and have plotted the
results with respect to the redshift. We have shown that the null
deviation vector fields and the observer area--distance, in the
model, have an evolution almost similar to the corresponding ones
in the $\Lambda$CDM model, the Hu--Sawicki models of $f(R)$ theory
and the $f(T)$ gravity.
\section*{Acknowledgments}
\indent

We thank the Research Office of Shahid Beheshti University for the
financial support.
\section*{Appendix}
\indent

From the definition of the interaction/coupling energy--momentum
tensor, Eq. (\ref{eq7}), one has
\begin{equation}\label{eq103}
\kappa {\nabla^\mu }T_{\mu \nu }^{[{\mathop{\rm int}} ]} = {\nabla
^\mu }\left[ {{f_T}T_{\mu \nu }^{[m]} + \frac{1}{2}\left( {f -
R{f_R}} \right){g_{\mu \nu }} + \left( {{\nabla _\mu }{\nabla _\nu
} - {g_{\mu \nu }}} \Box \right){f_R}} \right],
\end{equation}
wherein using the definition of Einstein tensor, it reads
\begin{equation}\label{eq105}
\kappa {\nabla^\mu }T_{\mu \nu }^{[{\mathop{\rm int}} ]} = {\nabla
^\mu }\left[ {{f_T}T_{\mu \nu }^{[m]} + \frac{f}{2}{g_{\mu \nu }}
+ {f_R}{G_{\mu \nu }} - {f_R}{R_{\mu \nu }} + \left( {{\nabla _\mu
}{\nabla _\nu } - {g_{\mu \nu }}} \Box \right){f_R}} \right].
\end{equation}
As for any scalar and vector fields, one has
${\nabla_\mu}{\nabla_\nu}\varphi
={\nabla_\nu}{\nabla_\mu}\varphi$\ \ and
\begin{equation}\label{eq106}
{\nabla_\mu }{\nabla _\nu }{A_\alpha } - {\nabla _\nu }{\nabla
_\mu }{A_\alpha } = {R_{\alpha \beta \mu \nu }}{A^\beta },
\end{equation}
thus, one gets
\begin{equation}\label{eq107}
{\nabla^\mu }{\nabla _\mu }{\nabla _\nu }{f_R} = {\nabla ^\mu
}{\nabla _\nu }{\nabla _\mu }{f_R} = {\nabla _\nu }{\nabla ^\mu
}{\nabla _\mu }{f_R} + {R_{\mu \alpha }}{^\mu _\nu }{\nabla
^\alpha }{f_R} = {\nabla _\nu }\square {f_R} + {R_{\mu \nu
}}{\nabla ^\mu }{f_R}.
\end{equation}
Also, by
\begin{equation}\label{eq108}
{\nabla_\nu }f\left( {R,T} \right) = {f_R}{\nabla _\nu }R +
{f_T}{\nabla _\nu }T
\end{equation}
and
\begin{equation}\label{eq109}
{\nabla^\mu }{G_{\mu \nu }} = 0\qquad \Rightarrow\qquad {\nabla
^\mu }{R_{\mu \nu }} = \frac{1}{2}{\nabla _\nu }R,
\end{equation}
one attains
\begin{equation}\label{eq110}
\frac{{{g_{\mu \nu }}}}{2}\left( {{\nabla^\mu }f} \right) -
{f_R}\left( {{\nabla ^\mu }{R_{\mu \nu }}} \right) =
\frac{1}{2}{f_T}{\nabla _\nu }T.
\end{equation}
Knowing ${\nabla_\mu }{g_{\alpha \beta }} = 0 $, and substituting
relations (\ref{eq107}) and (\ref{eq110}) into relation
(\ref{eq105}), leads to
\begin{equation}\label{eq111}
\kappa {\nabla^\mu }T_{\mu \nu }^{[{\mathop{\rm int}} ]} = T_{\mu
\nu }^{[m]}{\nabla ^\mu }{f_T} + {G_{\mu \nu }}{\nabla ^\mu }{f_R}
+ \frac{1}{2}{f_T}{\nabla _\nu }T
\end{equation}
wherein, with relation (\ref{eq conserv2}), it poses some
restrictions on the choice of the function of $f(R,T)$ on $T$ as
\begin{equation}\label{constrainEq}
T_{\mu \nu }^{[m]}{\nabla ^\mu }{f_T} + \frac{1}{2}{f_T}{\nabla
_\nu }T=0.
\end{equation}
Note that, such restrictions not only do~not contradict with the
diffeomorphism invariance of the action (\ref{eq1}), but they also
arise out of it. That is, the price for {\it a priori} inclusion
of the trace of the energy--momentum tensor (that itself will
emerge from the variation of the Lagrangian of the matter,
definition (\ref{eq2})) is to have these restrictions as
self--consistency.

Now, writing the result for the index $\nu=0 $ of the comoving
FLRW metrics, it yields
\begin{equation}\label{eq112}
-T_{00}^{[m]}{\dot f_T} + \frac{1}{2}{f_T}{\dot T^{[m]}} = 0,
\end{equation}
and by substituting $T_{00}^{[m]} = {\rho ^{[m]}}$ and ${\dot T} =
- {\dot \rho ^{[m]}} = 3H{\rho ^{[m]}} $, one achieves the
constraint equation (\ref{eq10}).

\end{document}